# Effects of Ion Milling on the Microwave Properties of MgB$_2$ Films

Sang Young Lee, J. H. Lee, J. Lim, H. N. Lee, S. H. Moon, B. Oh and M. A. Hein

*Abstract*— The new superconductor MgB$_2$ may prove useful for microwave applications at intermediate temperatures. MgB$_2$ films with the thickness of 300 – 400 nm have surface resistance $R_S$ less than 10 μΩ at a frequency (*f*) of 8.5 GHz and 7 K, and ~ 1.5 mΩ at 87 GHz and 4.2 K. The critical temperature ($T_C$) of these films is ~ 39 K when they are prepared under optimum conditions. The $R_S$ appears to scale as $f^2$ up to 30 K. After surface ion-milling, a reduction of the $T_C$ and an enhanced resistivity $\rho(T_C)$ are observed consistently at 8.5 GHz and 87 GHz along with a reduced $R_S$ at low temperatures. The observed $\rho(T_C)$ - $T_C$ relation and the uncorrelated behavior between $\rho(T_C)$ and $R_S$ values measured at low temperatures are well explained in terms of the two-gap model, with the observed $\rho(T_C)$ - $T_C$ relation attributed to the properties of the large gap, and the $R_S$ at lower temperatures reflecting the properties of the small gap, with an enhanced small gap energy due to increased interband scattering. This study suggests that the interband scattering should be enhanced to improve the low temperature microwave properties of MgB$_2$ films and that the ion-milling process must be performed with great care to preserve the high quality of MgB$_2$ films.

*Index Terms*— Ion-milling, MgB$_2$ film, Microwave properties, Two gap.

## I. Introduction

The boride superconductor MgB$_2$ discovered in early 2001 [1] appears attractive for device applications in an intermediate temperature range [2]; the critical temperature ($T_C$) is the highest among intermetallic compounds and the grain boundaries are strongly linked [3]. Unlike high-$T_c$ superconductors, MgB$_2$ is thought to have s-wave gap symmetry [4], which is expected to allow an exponential dependence of the surface impedance on temperature at low temperatures [5]-[7]. Furthermore, the fact that high-quality MgB$_2$ films can be prepared on sapphire substrate [8, 9] makes it easier to produce them at a relatively low cost for practical microwave applications.

Several factors must be understood and improved, however, before MgB$_2$ can become a attractive superconductor. These include the observed rapid drop of the critical current density under increased magnetic field [10] and the surface-sensitive character of MgB$_2$ films due possibly to the existence of a Mg-rich surface layer for films prepared by the two step process [11]. The possibility of two gaps in MgB$_2$, with the small gap apparently responsible for the $R_S$ of MgB$_2$ at low temperatures, also creates difficulty in optimizing the microwave properties of MgB$_2$ [12].

In this paper, we report effects of surface ion milling on the microwave properties of high-quality MgB$_2$ films, both in the normal state and in the superconducting state. The intrinsic surface resistance ($R_S$) of MgB$_2$ films prepared under optimized conditions was less than 10 μΩ at 8.5 GHz and 7 K, and ~ 1.5 mΩ at 87 GHz and 4.2 K, which was comparable to the corresponding typical $R_S$ values of epitaxially grown YBa$_2$Cu$_3$O$_{7-\delta}$ (YBCO) films [13]. The changes in the surface resistance due to ion-milling are well explained by the two-gap model for MgB$_2$.

## II. Experimental

The MgB$_2$ films were prepared on *c*-cut sapphire, MgO and LaAlO$_3$ by the two step process where boron films deposited on the substrates are annealed in a Mg vapor environment inside quartz tubes [9]. The films used for this experiment fall into three different groups. Films in group I (MB-1 to 4) were prepared in an earlier stage of this study and appeared to have relatively high $R_S$. The films in group II (MB-12 to 14) were surface ion-milled immediately after being prepared, and those in group III (MB-15 to 20) were prepared using an improved growth technique. In preparing the ion-milled films, the surface of the MgB$_2$ film was etched by argon ion-milling under an angle of 70 degree with respect to the film plane. The etching rate was ~10 nm/min with a beam voltage of 500 V and a current density of 0.28 mA/cm$^2$.

Properties of the MgB$_2$ films including the film thickness (*t*),

Manuscript received August 6, 2002.
This work was partially supported by the KOSEF under the grant No. R01-2001-00038, Korea Ministry of Science and Technology, MARC administered by Agency for Defense Development, and the National Research Laboratory Project

S. Y. Lee is with the National Institute of Standards and Technology, Boulder, CO 80305 USA, on leave from Konkuk University Seoul 143-701, Korea (phone: 303-497-5309; fax 303-497-3066; e-mail:sylee@boulder.nist.gov).
J. H. Lee is with Department of Physics, Konkuk University, Seoul 143-701, Korea (e-mail: jaju@konkuk.ac.kr).
J. Lim is with Department of Physics, Konkuk University, Seoul 143-701, Korea (e-mail: jjun@konkuk.ac.kr).
H. N. Lee is with LG Electronics Institute of Technology, Seoul 137-724, Korea (e-mail: hnlee@lge.com).
S. H. Moon is with Seoul National University, Seoul, Korea (e-mail: smoon@gong.snu.ac.kr).
B. Oh is with LG Electronics Institute of Technology, Seoul 137-724, Korea (e-mail: boh@lge.com).
M. A. Hein is with University of Wuppertal, Wuppertal, Germany (e-mail: mhein@venus.physik.uni-wuppertal.de)

the onset critical temperature determined at high frequency $T_C$(HF), the normal-state surface resistance ($R_S^N$), and the normal-state resistivity $\rho(T_C)$(HF) calculated from the normal-state surface resistance at $T_C$(HF) are listed in Table I.

Microwave properties were measured at ~ 8.5 GHz and ~ 87 GHz using a $TE_{01\delta}$ mode rutile-loaded resonator and a $TE_{01\delta}$ mode hollow cavity resonator at temperatures between 4.2 K and 45 K. The temperature was stable within ± 0.15 K. The effective surface resistance $R_S^{eff}$ was obtained from the unloaded quality factor measured in a weak-coupling scheme. The measured $R_S^{eff}$ was reproducible within 5 % with errors in $R_S^{eff}$ at 8.5 GHz up to ± 30 % below 10 K due to errors in the measured loss tangent of rutile. More details on the measurement procedures are described elsewhere [11, 13]. The intrinsic surface resistance $R_S$ was calculated using both the impedance transformation method [14] and a method based on a rigorous $TE_{01\delta}$ mode analysis [15]. No significant difference was observed for these two methods for films with $t/\lambda > 1$ and the relation $R_S^{eff} = \{\coth(t/\lambda) + (t/\lambda)/\sinh^2((t/\lambda))\} \times R_S$ was used for this purpose. The normal-state intrinsic surface resistance $R_S^N$ was also calculated from the measured normal-state effective values ($R_S^{N,eff}$) with the finite thickness of the films taken into account and assuming normal skin effect for $MgB_2$ in the normal state. The normal-state resistivity $\rho(T_C)$(HF) was obtained at the critical temperature $T_C$(HF) using $R_S^N$. $T_C$(HF) appeared to depend on the measurement frequency.

## III. RESULTS AND DISCUSSION

Figure 1 shows the temperature dependence of the effective surface resistance ($R_S^{eff}$) for MB 4A, 12II, 19A and 20A. The $R_S^{eff}$ of MB 12II and 20A are very low at low temperatures, with $R_S^{eff}$(10 K) ~ 3.6 mΩ and ~ 3.4 mΩ at 87 GHz, respectively. Assuming a penetration depth of ~ 160 nm, the corresponding intrinsic surface resistance $R_S$ would be 3.1 mΩ and 3.2 mΩ at 87 GHz, respectively, which correspond to the values of ~ 40 μΩ when scaled to 10 GHz using the $R_S \sim f^2$ relation. In the figure, the $R_S^{eff}$ value of MB 4A is about an order of magnitude higher than that of MB 12II and 20A at ~ 25 K although its $R_S^{N,eff}$ value is significantly lower. It is noted in Fig. 1 that the $R_S^{eff}$ values at 8.5 GHz and 87 GHz follow the $R_S \sim f^2$ relation up to temperatures of ~ 30 K for MB12-II, 19A and 20A as can be explained by the two-fluid model. To clarify this, we calculated the $R_S$ of $MgB_2$ films by taking the finite thickness of the films into consideration. For the calculation, the values of the penetration depth ($\lambda$) were obtained from the changes in the resonant frequency of the resonator using a simple BCS model. Figure 2 and its inset shows the $R_S$ vs. frequency data for MB 18A and MB 20A, where the $R_S$ values are seen to follow the $R_S \sim f^2$ relation up to 30 K.

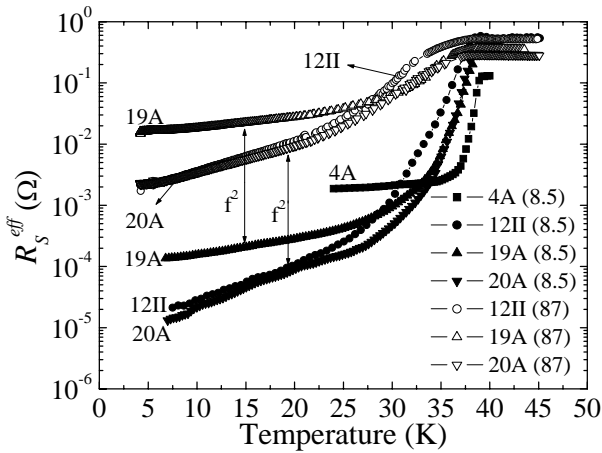

Fig. 1. The temperature dependence of $R_S^{eff}$ measured at 8.5 GHz (filled symbols) and 87 GHz (open symbols) for MB 4A, 12II, 19A, 19I, 20I and 20II. The $R_S^{eff}$ of 12II, 19A and 20A is seen to scale as $f^2$ at temperatures up to ~ 30 K.

TABLE I.
SAMPLE NUMBERS, VALUES OF $T_C$ AND $\rho(T_C)$ MEASURED AT 8.5 GHZ AND 87 GHZ FOR ALL THE $MgB_2$ FILMS, AND $R_S^N$ VALUES AT 8.5 GHZ. IN LABELING THE SAMPLES, 'A', 'I' AND 'II' ARE USED TO DENOTE 'AS-PREPARED', 'ONCE ION-MILLED' AND 'THE SECOND ION-MILLED', RESPECTIVELY. THE THICKNESS DIFFERENCE BETWEEN THE SAMPLES STARTING WITH THE SAME NUMBER (E.G. 1A AND 1I) INDICATES THE FILM THICKNESS REMOVED BY ION-MILLING. THE KINDS OF SUBSTRATE MATERIALS USED FOR THE GROWTH OF ALL THE FILMS ARE ALSO LISTED. IN THE TABLE, 'SAPPHIRE', 'MgO' AND 'LaAlO$_3$' MEANS $C$-CUT SAPPHIRE, (100)MgO AND (100)LaAlO$_3$, RESPECTIVELY.

| Sample No. | Substrate | Thick-ness (nm) | $T_C$(HF) at 8.5GHz (K) | $T_C$ at 87GHz (K) | $R_S^N(T_C)$ at 8.5GHz (Ω) | $\rho(T_C)$ at 8.5GHz (μΩ-cm) | $\rho(T_C)$ at 87GHz (μΩ-cm) |
|---|---|---|---|---|---|---|---|
| MB-1A | Sapphire | 420 | 38.9 | | 0.068 | 13.8 | |
| 1I | | 365 | 39.1 | | 0.059 | 10.4 | |
| 2A | Sapphire | 420 | 36.9 | | 0.034 | 3.4 | |
| 2I | | 365 | 37.6 | | 0.032 | 3.1 | |
| 2II | | 320 | 36.9 | | 0.042 | 5.3 | |
| 3A | MgO | 420 | 37.8 | | 0.048 | 6.9 | |
| 3II | | 320 | 37.3 | | | | |
| 4A | MgO | 400 | 39.2 | | 0.042 | 5.3 | |
| 4I | | 350 | 39.1 | | 0.045 | 6.0 | |
| 4II | | 325 | 37.0 | | 0.043 | 5.4 | |
| 11A | Sapphire | 420 | 38.0 | | 0.068 | 13.8 | |
| 12I | Sapphire | 370 | 39.6 | | 0.075 | 16.6 | |
| 12II | | 345 | 39.1 | 37.8 | 0.082 | 20.1 | 17.9 |
| 13I | Sapphire | 380 | 37.3 | 37.8 | 0.070 | 14.4 | 18.0 |
| 13II | | 310 | 36.0 | 31.9 | 0.066 | 12.8 | 27.1 |
| 14I | LaAlO$_3$ | 380 | 37.2 | 37.9 | 0.073 | 15.9 | 17.0 |
| 14II | | 310 | 36.5 | 35.0 | 0.069 | 14.0 | 17.0 |
| 15A | Sapphire | 370 | 39.2 | | 0.069 | 14.0 | |
| 16A | Sapphire | 370 | 39.0 | | 0.070 | 14.6 | |
| 16I | | 310 | 38.5 | | 0.071 | 15.2 | |
| 17A | Sapphire | 370 | 39.0 | | 0.068 | 13.6 | |
| 17I | | 310 | 37.9 | | 0.070 | 14.8 | |
| 18A | Sapphire | 330 | 38.8 | 37.8 | 0.067 | 13.3 | 12.3 |
| 19A | Sapphire | 330 | 39.5 | 38.2 | 0.064 | 12.1 | 11.8 |
| 19I | | 270 | 38.0 | 36.9 | 0.067 | 13.3 | 13.7 |
| 20A | MgO | 450 | 38.9 | 37.7 | 0.082 | 20.0 | 12.8 |
| 20I | | 390 | 36.3 | 35.4 | 0.082 | 20.3 | 20.5 |

Changes in the microwave properties were significant for all the $MgB_2$ films after surface ion-milling. First, the critical temperature of all the as-prepared $MgB_2$ films in group III decreased after surface ion-milling as determined from the temperature-dependent $R_S^{eff}$ data. Figure 3(a) shows the $T_C$(HF) vs. $\rho(T_C)$(HF) data for as-prepared films MB15A-20A and ion-milled films MB 16I - 20I. In all cases, the decrease in $T_C$(HF) is accompanied by an increase in $\rho(T_C)$(HF) after ion-milling. For most films, the ratio of $\Delta T_C$(HF) to $\rho(T_C)$(HF) is ~ 1 K/μΩ-cm. The increase of $\rho(T_C)$(HF) after ion milling is very likely due to the increased defect density created during the ion milling process and the resulting increase of the scattering rate of electrons. We note that a similar correlation between $T_C$ and $\rho$ ($T_C$) has been observed for A15 compounds and bcc transition metals, which is attributed to electron lifetime effects by Testardi and Mattheiss [16]. Although $\rho(T_C)$(HF) and $\rho_0$ cannot be directly compared with each other, it is very likely that increased defect density in MgB$_2$ films would result in an increase in $\rho_0$ as well. For reference, $\Delta T_C/\rho_0$ is 0.1 – 0.2 K/μΩ-cm for A15 compounds and bcc transition metals with low $\rho_0$.

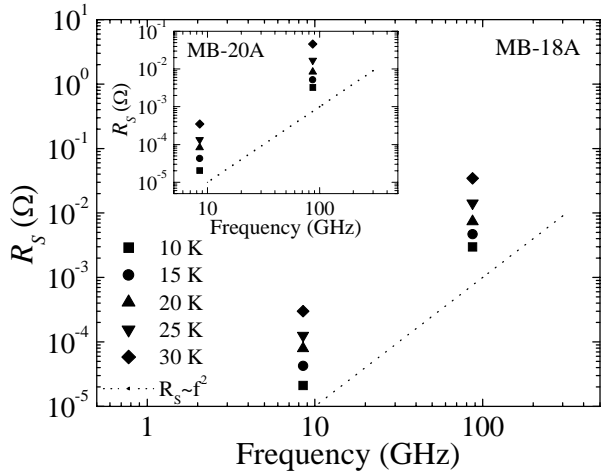

Fig. 2. Frequency dependence of the intrinsic surface resistance $R_S$ between 10 K and 30 K for MB 18A. The dotted line is a guide showing the quadratic frequency dependence. The slope appears slightly more than 2 for the $R_S$ at 10 K due to measurement errors in obtaining the loss tangent of rutile.
Inset: Frequency dependence of the intrinsic surface resistance $R_S$ between 10 K and 30 K for MB 20A. The dotted line is a guide showing the quadratic frequency dependence.

When the MgB$_2$ films were ion-milled twice, the correlation between $T_C$(HF) and $\rho(T_C)$(HF) was not clear. In Fig. 3(b), we show the data for MB 11A, MB 12I-14I (films ion-milled one time) and MB 12II-14II (films ion-milled twice) measured at 8.5 GHz and 87 GHz. The data at 87 GHz still show some correlation between $T_C$(HF) and $\rho(T_C)$(HF) for MB 13 and MB 14. However, at 8.5 K, both $T_C$(HF) and $\rho(T_C)$(HF) decrease after ion-milling. The difference in the $T_C$(HF) vs. $\rho(T_C)$(HF) behavior for MB 13 and MB 14 between the data at 8.5 GHz and 87 GHz is not understood at this time.

In Figs. 3(a) and 3(b), differences in $T_C$(HF) and $\rho(T_C)$(HF) are seen between the data at 8.5 GHz and 87 GHz. For the films in groups II and III, the measured $T_C$(HF) values at 87 K are consistently lower than the corresponding ones at 8.5 GHz. Within the context of the two-gap scenario for MgB$_2$, the observed difference in the $T_C$(HF) values may be related to the presence of the small energy gap, since 87 GHz is about 10 % of the gap frequency for $\Delta/k_BT_C \sim 1$. There are two exceptions to this general observation regarding $T_C$(HF); the $T_C$(HF) values of MB13I and MB14I at 87 GHz are a little higher than the corresponding ones at 8.5 GHz, respectively. However, the observed small differences in $T_C$(HF) for MB13I and 14I may be within the measurement errors for determining $T_C$(HF).

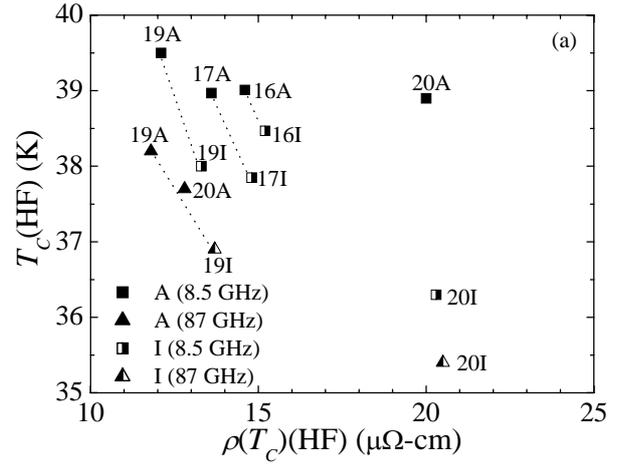

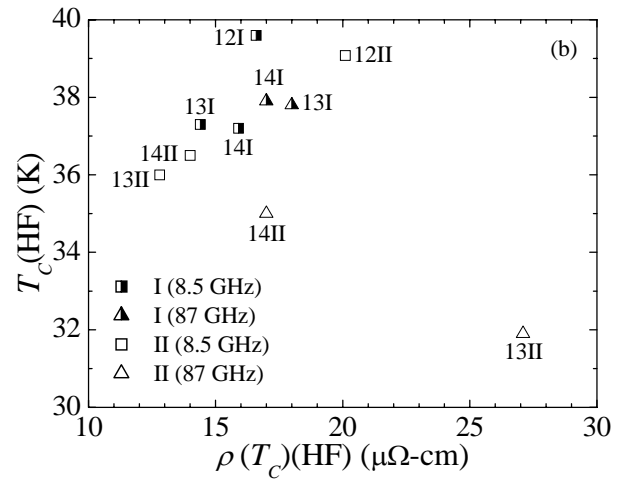

Fig. 3(a). $\rho(T_C)$(HF) vs. $T_C$(HF) data at 8.5 GHz and 87 GHz for MgB$_2$ films in group III (MB 16 - 20). All the films show reduction in $\rho(T_C)$(HF) and $T_C$(HF) after ion-milling. The dotted lines are guides to eyes showing typical changes in $\rho(T_C)$(HF) and $T_C$(HF) due to ion-milling for films in group III.
Fig. 3(b). $\rho(T_C)$(HF) vs. $T_C$(HF) data at 8.5 GHz and 87 GHz for MgB$_2$ films in group II (MB 12 – 14). Unlike the films in group III (MB 16 – 20), the films once ion-milled do not show consistent reduction in $\rho(T_C)$(HF) and $T_C$(HF) at 8.5 GHz after a second ion-milling.

In Figs. 3(a) and 3(b), it is seen that the $\rho(T_C)$(HF) of the films in groups II and III is in the range of 10 - 20 μΩ-cm regardless of the ion-milling process. In particular, the $\rho(T_C)$(HF) values of the as-prepared films with the lowest $R_S^{eff}$ values (e.g., MB 15A, 16A, 18A and 20A) are in the range of 12 – 15 μΩ-cm. These values are still significantly higher than the values of $1 - 5$ μΩ-cm, which have been frequently reported for single crystal MgB$_2$ and bulk polycrystalline MgB$_2$. The difference in the $\rho(T_C)$(HF) values at 8.5 GHz and 87 GHz can be understood in terms of the measurement errors. In most cases except for MB-13II, the differences in the measured $\rho(T_C)$(HF) values at 8.5 GHz and 87 GHz are within 10 %, which can be explained assuming an measurement error of 5 % near $T_C$(HF), a plausible number considering that the resonator coupling becomes drastically weak near $T_C$(HF).

Very interesting features were observed in the temperature-dependent $R_S^{eff}$ data for the as-prepared and ion-milled MgB$_2$ films. Figure 4 shows the $R_S^{eff}$ vs. the reduced temperature ($T/T_C$) data for MB-19A, 19I, 20A and 20I measured at 8.5 GHz and 87 GHz. In the figure, we see a crossover in the $R_S^{eff}$ vs. $T/T_C$(HF) curves for the as-grown and the ion-milled MgB$_2$ films with the $R_S^{eff}$ of the ion-milled films becoming smaller than that of the as-grown films despite the degradation of $T_C$(HF) after ion-milling. It is noted that the crossover temperature appears almost the same at 8.5 GHz and 87 GHz, occurring at ~ 0.85 for MB 19A and 19I, and ~ 0.6 for MB 20A and 20I. Such a crossover can be well understood within the context of the two-gap scenario along with the reduction in $T_C$(HF). First, $T_C$(HF) cannot be affected by intraband scattering according to Anderson's theorem. In this regard, it is believed that the reduction in the $T_C$(HF) of the as-grown MgB$_2$ after ion-milling is attributed to the increased interband scattering between the σ-band and the π-band in MgB$_2$. Since the $R_S$ values would be dominated by the properties of the small gap at low temperatures, the reduced $R_S$ of the ion-milled MgB$_2$ at low temperatures would mean an increased gap energy of the small gap due to the increased interband scattering by ion milling-induced defects. No correlation was found between $\rho(T_C)$(HF) and the $R_S$ measured at low temperatures. Within this model, the $\rho(T_C)$(HF)- $T_C$(HF) relation is attributed to the properties of the large gap, i.e, the σ-band, with the $R_S$ at lower temperatures reflecting the properties of the small gap, i.e., the π-band when the uncorrelated $\rho(T_C)$(HF)-$R_S$ behavior is considered.

Our arguments are consistent with the recent report by Mazin et al. [17]. According to them, although the two-band model predicts lower $T_C$ for MgB$_2$ samples with higher $\rho_0$, the extremely small interband scattering in MgB$_2$ enables many MgB2 samples to have similar $T_C$ despite having a distribution in $\rho_0$. If MgB$_2$ samples can be made with enhanced interband scattering, lower $T_C$ should correlate with higher $\rho_0$. We therefore attribute the reduced $T_C$(HF) we observe due to ion-milling as due to enhanced interband scattering.

It is also noted that for MgB$_2$ films, repeated ion-milling does not always yield reduced $R_S$ at low temperatures. In the inset of Fig. 4, the $R_S^{eff}$ vs. $T/T_C$ data for MB 12I and 12 II show reduction in the $T_C$ and $R_S$ of MB12 I after ion-millng as observed for films in group III. Similar behavior was observed for MB-4A and 4I. However, when MB-4I was ion-milled again, its $R_S^{eff}$ appeared significantly enhanced, with a reduction in $R_S^N$ (see the inset of Fig. 4). This suggests that reduction in $R_S$ at low temperatures only be realized when modest disorder is introduced in MgB$_2$. Similar observations have been made in proton-irradiated bulk MgB$_2$ by Bugoslavsky et al. [18], who reported enhancement of the high-magnetic field critical current density of bulk MgB$_2$ by proton irradiation. In this regard, excessive disorder might have played some roles in enhancement of $R_S$ for ion-milled MgB$_2$ films [5] and reduction

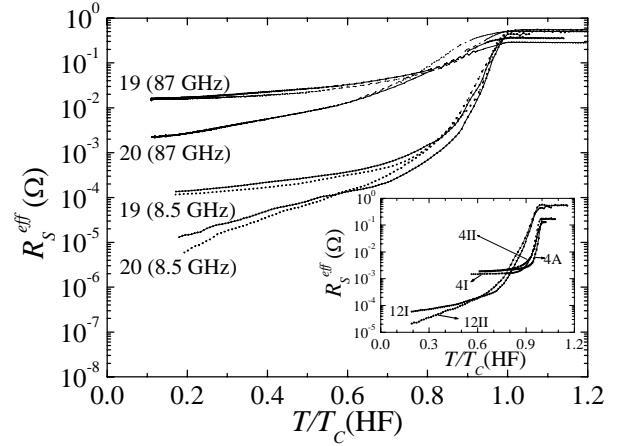

Fig. 4. Dependence of $R_S^{eff}$ on the reduced temperature $T/T_C$(HF) for MB-19A, 19I, 20A, 20I at 8.5 GHz and 87 GHz. Crossovers are seen between the $R_S^{eff}$ values of the as-prepared films and the once ion-milled films with the crossover temperature appearing nearly the same for each pair of films at 8.5 GHz and 87 GHz. The $R_S^{eff}$ of MB-20A and 20I looks almost the same for $T/T_C$(HF) < 0.6. However, the intrinsic $R_S$ of MB-20I is smaller than that of MB-20A after the corrections for film thickness.
Inset: Dependence of $R_S^{eff}$ on the reduced temperature $T/T_C$(HF) for MB-4A, 4I, 4II, 12I and 12II at 8.5 GHz. The $R_S^{eff}$ of MB-4II appear significantly enhanced compare to that of MB-4I.

of the critical current density for electron-irradiated bulk MgB$_2$ at high magnetic field [19].

For films not prepared under optimized conditions, a Mg-rich metallic layer could affect their $R_S$ values significantly [11]. Such films usually show significantly lower $R_S^N$ (see e.g., Table I for the $R_S^N$ values for MB 2 - 4 in group I) compared to the values observed for films in group II and III. For these films, $T_C$ also appeared to change little after the ion-milling.

A similar crossover behavior has been reported in the $R_S$ vs. temperature curves for YBCO films with different defects densities [20], with the crossover attributed to the behavior of $\sigma_1$, the real part of complex conductivity, due to the temperature-dependent inelastic scattering rate and the quasiparticle density. To date, however, the $\rho(T_C)$(HF)-$T_C$(HF) relation and the $\rho(T_C)$(HF)-$R_S$ relation are not known for YBCO films, which makes it difficult to compare the properties of YBCO with those of MgB$_2$.

Our interpretations of the changes in the properties of MgB$_2$ films are also consistent with our recent observations for the nonlinear behavior of as-prepared and ion-milled films. At relatively high temperatures, the ion-milled films showed higher nonlinear response. However, as the temperature decreases, a crossover was observed between the nonlinear response of the as-prepared and the ion-milled films, resulting in the observation of lower nonlinear response from the ion-milled films. This observation also fits the model of the small gap becoming enhanced due to the ion-milling induced interband scattering. More details on the nonlinear responses of MgB$_2$ films will be reported elsewhere [21].

## IV. Conclusion

High-quality $MgB_2$ films with $R_S$ less than 10 μΩ at 8.5 GHz and 7 K, and ~ 1.5 mΩ at 87 GHz and 4.2 K, with the critical temperature ($T_C$) of ~ 39 K have been prepared, and effect of ion-milling on their microwave properties were studied. After ion-milling, reduction of the $T_C$ and enhancement of the normal-state resistivity were consistently observed along with a reduction of $R_S$ at low temperatures. The intrinsic $R_S$ appeared to scale with $f^2$ up to 30 K. The observed $\rho(T_C)$(HF) - $T_C$(HF) relation and the uncorrelated behavior between $\rho(T_C)$(HF) and $R_S$ measured at low temperatures are well explained within the context of the two-gap model, with the observed $\rho(T_C)$(HF) - $T_C$(HF) relation attributed to the properties of the large gap, i.e., the $\sigma$-band, and the $R_S$ at lower temperatures reflecting the properties of the small gap, i.e., the $\pi$-band, with the small gap energy enhanced by increased interband scattering. In this regard, finding ways to increase the interband scattering seems to be one of the key means to improve the low temperature microwave properties of $MgB_2$, which is dominated by the small gap.

Our results also show that ion-milling, which is frequently used for passivation or planarization for thin films, affects the superconducting properties of $MgB_2$ films significantly and great care needs to be taken for the properties of $MgB_2$ films to be preserved when any processes involving ion-milling are performed.

## Acknowledgment

Valuable support from James C. Booth and David Rudman at NIST, H. J. Lee at POSTECH and S.-G. Lee at Korea University is greatly acknowledged.

## References

[1] J. Nagamatsu, N. Nakagawa, T. Muranaka, Y. Zenitani, and J. Akimitsu, "Superconductivity at 39 K in $MgB_2$," *Nature*, vol. 410, pp. 63-64, Mar. 2001.

[2] See e.g., M. A. Hein, "Perspectives of superconducting $MgB_2$ for microwave applications", presented at URSI-GA, Maastricht, Germany, Aug. 2, 2002.

[3] D. C. Labalestier et al., "Strongly linked current flow in polycrystalline forms of the new superconductor $MgB_2$," *Nature*, vol. 410, pp. 186-189, Mar. 2001.

[4] See e. g., C. Buzea and T. Yamashita, "Review of the superconducting properties of $MgB_2$," *Supercond. Sci. Technol.*, vol. 14, R115-R146, Nov. 2001, and references therein.

[5] A. A. Zhukov et al., "Temperature dependence of the microwave surface impedance measured on different kinds of $MgB_2$ films," unpublished

[6] A. Andreone et al., "Electrodynamic response of $MgB_2$ sintered pellets and thin films, "in *Studies of High Temperature Superconductors*, vol. 41, A.V. Narlikar, Ed. New York: Nova Sci. Publ., to be published.

[7] B. B. Jin, N. Klein, W. N. Kang, H.-J. Kim, E.-M. Choi, and S.-I. Lee, "Energy gap, penetration depth and surface resistance of $MgB_2$ films determined by microwave resonator measurements," submitted for publication.

[8] W. N. Kang, H. J. Choi, E. M. Kim, K. H. P. Kim and S. I. Lee, "$MgB_2$ superconducting thin films with a transition temperature of 39 Kelvin," *Science*, vol. 292, pp. 1521-1522, Apr. 2001.

[9] S. H. Moon, J. H. Yun, H. N. Lee, J. I. Kye, H. G. Kim, W. Chung, and B. Oh, "High critical current densities in superconducting $MgB_2$ films," *Appl. Phys. Lett.*, vol. 79, pp. 2429-2431, Oct. 2001.

[10] P. Vase, R. Flukiger, M. Leghissa, and B. A. Glowacki, "Current status of high-$T_C$ wire," *Supercond. Sci. Technol.*, vol. 13, pp. R71-R84, Jul. 2000.

[11] Sang Young Lee, J. H. Lee, Jung Hun Lee, J. S. Ryu, and J. Lim, "Significant reduction of the microwave surface resistance of $MgB_2$ films by surface ion milling," *Appl. Phys. Lett.*, vol. 79, pp. 3299-3301, Nov. 2001.

[12] A. A. Golubov, A. Brinkman, O. V. Dolgov, J. Kortus, and O. Jepsen, "Multiband model for penetration depth in $MgB_2$," unpublished; H. Schmidt, J. F. Zasadzinski, K. E. Gray, and D. G. Hinks, "Energy gap from tunneling and metallic contacts onto $MgB_2$: Possible evidence for a weakened surface layer," *Phys. Rev. B*, vol 63, 220504-1 – 220504-3, May 2001; Mun-Seog Kim, J. A. Skinta, T. R. Lemberger, "Reflection of two-gap nature in penetration depth measurements of $MgB_2$ films," unpublished.

[13] See e.g., M. Hein, *High-Temperature Superconductor Thin Films at Microwave Frequencies*, Heidelberg, Germany: Springer Tracts of Modern Physics, 155. Springer, 1999, Ch. 2; See also M. J. Lancaster, *Passive Microwave Device Applications of High-Temperature Superconductors*, Cambridge, UK, Cambridge University Press, 1997.

[14] N. Klein et al., "The effective microwave surface impedance of high-$T_C$ thin films," *J. Appl. Phys.*, vol. 67, pp. 6940-6945, Jun. 1990.

[15] Sang Young Lee, J. H. Lee, J. C. Booth and V. B. Fedorov, "Theoretical analysis for the effective surface impedance of superconductor films used for a $TE_{011}$ mode dielectric-loaded cavity resonator," unpublished.

[16] L. R. Testardi and L. F. Mattheiss, "Electron lifetime effects on properties of A15 and bcc metals," *Phys. Rev. Lett.*, vol. 41, pp. 1612-1615, Dec. 1978.

[17] I. I. Mazin, O. K. Andersen, O. Jepsen, O. V. Dolgov, J. Kortus, A. A. Golubov, A. B. Kuz'menko, and D. van der Marel, "Two-gap superconductivity in $MgB_2$: clean or dirty?," unpublished.

[18] Y. Bugoslavsky, L. F. Cohen, G. K. Perkins, M. Pollchetti, Y. J. Tate, R. Gwilliam, and A. D. Caplin, Enhancement of the high-magnetic field critical current density of superconducting $MgB_2$ by proton irradiation," *Nature*, vol. 411, pp. 561-563, May 2001.

[19] S. Okayasu, H. Iketa, and R. Yoshizaki, "Electron irradiation effects on $MgB_2$ bulk samples," unpublished.

[20] J. Einfeld, P. Lahl, R. Kutzner, R. Wordenweber, G. Kastner, "Reduction of the microwave surface resistance in YBCO thin films by microscopic defects," *Physica C*, vol. 351, pp. 103-117, 2001.

[21] J. C. Booth, Sang Young Lee, K. Leong, J. H. Lee, J. Lim, H. N. Lee, S. H. Moon and B. Oh, "Nonlinear properties and surface impedance of $MgB_2$ films," submitted for publication.